\documentclass[%
 reprint,
 amsmath,amssymb,
 aps,
]{revtex4-2}

\usepackage{graphicx}
\usepackage{dcolumn}
\usepackage{bm}
\usepackage{xcolor}

\begin{document}

\preprint{APS/123-QED}

\title{Logarithmic vs Andrade's transient creep: the role of elastic stress redistribution}

\author{Jérôme Weiss}
    \email{jerome.weiss@univ-grenoble-alpes.fr}
\author{David Amitrano}%
 
\affiliation{%
 Univ. Grenoble Alpes, CNRS, ISTerre, 38000 Grenoble, France
}%

\date{\today}

\begin{abstract}
Creep is defined as time-dependent deformation and rupture processes taking place within a material subjected to a constant applied stress smaller than its athermal, time-independent strength. This time-dependence is classically attributed to thermal activation of local deformation events. The phenomenology of creep is characterized by several ubiquitous but empirical rheological and scaling laws. We focus here on primary creep following the onset of loading, for which a power law decay of the strain-rate is observed, $\dot{\varepsilon}\sim t^{-p}$, with the exponent $p$ varying between $\simeq$0.4 and 1, this upper bound defining the so-called logarithmic creep. Although this phenomenology is known for more than a century, the physical origin of Andrade-like ($p<$1) creep remains unclear and debated. Here we show that $p<$1 values arise from the interplay between thermal activation and elastic stress redistribution. The latter stimulates creep dynamics from a shortening of waiting times between successive events, is associated to material damage and possibly, at high temperature and/or stresses, gives rise to avalanches of deformation events. 
\end{abstract}

\maketitle


\section{INTRODUCTION}

Creep and associated time-dependent deformation and rupture processes under an applied constant stress are of tremendous importance in various fields, from metallurgical engineering \cite{kassner2015fundamentals} to civil engineering \cite{bazant1995creep}, rocks mechanics \cite{brantut2012micromechanics} and geophysics \cite{savage2005postseismic}, or soft matter physics \cite{cipelletti2020microscopic}. The phenomenology of creep under constant load sums up, from the onset of loading, to a decelerating primary or transient creep, followed by a stage of constant strain-rate $\dot{\varepsilon}_{min}$, and ending by an accelerating tertiary creep leading to either a macroscopic rupture for brittle material, or fluidization in soft matter \cite{siebenburger2012creep}. Note that, in many materials, the secondary creep actually resumes to an inflexion point between primary and tertiary creep. This phenomenology is characterized by several empirical, but surprisingly universal, rheological and scaling laws, which however still lack nowadays a sound understanding. Examples are e.g. the Monkman-Grant expression relating the creep failure time $t_f$ to $\dot{\varepsilon}_{min}$ \cite{monkman1956empirical}, or a finite-time singularity characterizing tertiary creep, $\dot{\varepsilon} \sim (t_f-t)^{-\beta}$ \cite{koivisto2016predicting,leocmach2014creep}.   

We will focus here on primary creep, characterized by a power law decay of the strain rate, $\dot{\varepsilon}\sim(c+t)^{-p}$, where $c$ is a small delay timescale sometimes difficult to estimate experimentally. This phenomenology is shared by various materials and media, including metals \cite{andrade1910viscous}, rocks \cite{scholz1968mechanism,heap2009time}, ice \cite{schulson2009creep}, concrete \cite{vandamme2009nanogranular}, paper \cite{koivisto2016predicting}, colloidal glasses \cite{siebenburger2012creep}, gels \cite{leocmach2014creep,cho2021yield}, or dry granular media \cite{amon2012hot,miksic2013evolution}. This universality is surprising, owing to the various microscopic mechanisms involved in the creep deformation of these different materials: dislocations motion and interactions in metals, microcracking events in brittle rocks and concrete, shear transformation events in amorphous materials, or other mechanisms in biological and soft matter. This suggest some underlying common physics which remains partly obscure nowadays.  

In rocks, the following empirical temperature- and stress-dependence was proposed\cite{carter1978transient},
\begin{equation} \label{eq:Carter}
    \dot{\varepsilon}=B\sigma^{n}\text{exp}\left(-\frac{E}{k_BT}\right)t^{-p}
\end{equation}
, where $B$ is a material constant, $E$ a material-dependent but stress-independent activation energy, and $1\lesssim n\lesssim2$ a creep exponent.
In his historical paper, Andrade reported for metal wires an exponent $p$=2/3 \cite{andrade1910viscous}, henceforth defining the so-called Andrade's creep. On the other hand, a $p$=1 value is often reported, signing the logarithmic transient creep (as $\varepsilon$ grows as log$(c+t)$). As a matter of fact, if $p$=1 appears as an upper bound, reported experimental values actually range between $\sim$0.4 and 1 \cite{carter1978transient}. At least in metals and rocks, the $p$-value is observed to decrease with increasing imposed stress and/or temperature \cite{carter1978transient,cottrell1952time}, in contradiction with eq. (\ref{eq:Carter}), while in soft matter $p$ seems to increase with increasing aging \cite{siebenburger2012creep}. So, the $p$-value is not material-specific, while the physical origin of Andrade-like ($p<$1) creep remains unclear and debated (e.g. \cite{cottrell2004microscopic,louchet2009andrade,cho2021yield,miguel2002dislocation}).

A logarithmic increase of strain under constant stress (creep test), $\varepsilon\sim\text{log}(c+t)$, can be mirrored by a logarithmic decrease of the stress under constant strain (relaxation test), $\sigma\sim\text{-log}(c+t)$. This suggests interesting links with entirely different systems exhibiting slow, logarithmic relaxations, such as conductance relaxation in electron glasses \cite{amir2012relaxations} or the magnetic relaxation in superconductors \cite{kodama1997relation}. However, as shown below, there is no equivalent of Andrade's creep ($p<1$) under relaxation conditions.

Here we show, first inspired by a seminal work of Cottrell on dislocation-driven creep in metals \cite{cottrell1952time}, that logarithmic creep represents an upper bound for $p$, resulting from a thermally-activated exhaustion of weak spots, while ignoring elastic interactions and stress redistribution. Taking into account the increase of the effective stress acting on the remaining spots, the softening (damage) of the material, as well as the triggering of avalanches of deformation events from elastic interactions, leads to a decrease of the apparent $p$-value, but the $p$=2/3 historical value does not emerge as a special one. Instead, $p$ decreases with increasing imposed load, temperature, or disorder strength, in agreement with observations.

\section{A simple exhaustion model without stress redistribution} 

We consider first the simplest situation where only disorder and thermally-activated exhaustion combine. The following reasoning is an adaptation of the original Cottrell’s model \cite{cottrell1952time}, initially proposed to explain dislocation-driven logarithmic creep in metals. It is presented here to show what can be expected under this strong simplification, as well as its shortcomings. 
A system is assumed to be composed of $N$ elementary volumes with uniformly distributed activation energies $E_i$, $i=\{1,..,N\}$. A simple choice \cite{cottrell1952time,castellanos2018avalanche,cho2021yield} is to set $E_i=V_a\Delta\sigma_i$, where $V_a$ is a constant activation volume and $\Delta\sigma_i=\sigma_i-\sigma$ the stress gap between the local stress threshold $\sigma_i$ and the applied stress $\sigma$. Non-linear dependencies of $E$ on $\Delta\sigma$ have been considered as well, depending on the structural process involved \cite{mott1948report,merabia2016thermally,chattoraj2010universal}. The probability per unit time for an element with a stress gap $\Delta\sigma$ to be thermally activated is:
\begin{equation} \label{eq:thermal}
    P(\sigma,T)=\nu_0 \text{exp}\left(-\frac{V_a\Delta\sigma}{k_BT}\right)
\end{equation}
, where $\nu_0$ is an attempt frequency, $k_B$ the Boltzmann's constant and $T$ the temperature. Each thermally activated element contributes for a (fixed) strain increment $\delta\epsilon$ (corresponding to either an elementary dislocation motion, a microcracking event, a shear transformation,..), and is then exhausted, i.e. cannot be activated again. The crudest assumption of this model is that each deformation event occurs independently from the previous ones. This means that neither a damage of the material, nor the triggering of athermal deformation avalanches are considered. In such simple model, upon imposing the stress $\sigma$ at $t$=0, an athermal deformation can immediately occur from the triggering of the elements with $\sigma_i\leq\sigma$. Creep deformation ensues as a succession of independent thermally activated events, each contributing a strain increment $\delta\epsilon$, and the elements associated to a smaller stress gap $\Delta\sigma_i$ having, from (\ref{eq:thermal}), a greater chance to occur first before to be ruled out. Consequently, the inter-event time $\Delta t_i\sim1/\nu_i$ grows in average with time, so the creep strain-rate $\dot{\varepsilon}$ decreases. 
In this framework, the strain-rate is given by:
\begin{equation} \label{eq:Cottrell1}
    \dot{\varepsilon}=\delta\epsilon \int_{0}^{1} n(\Delta\sigma,t)P(\sigma) \,d{\Delta\sigma} 
\end{equation}
, where $n(\Delta\sigma,t)d{\Delta\sigma}$ is, at time $t$, the number of elements with a stress gap within $[\Delta\sigma,\Delta\sigma+d{\Delta\sigma}]$, normalized by $N$.

The exhaustion assumption can be written as:
\begin{equation} \label{eq:Cottrell2}
    \frac{\partial n(\Delta\sigma,t)}{\partial t}=-n(\Delta\sigma,t)P(\Delta\sigma)
\end{equation}
, which, after integration, gives:
\begin{equation} \label{eq:Cottrell3}
   n(\Delta\sigma,t)=n(\Delta\sigma,t=0) \text{exp}(-P(\Delta\sigma)t)
\end{equation}
As the stress gaps $\Delta\sigma$ are uniformly distributed between 0 and 1, $n(\Delta\sigma,t=0)$ is a constant.

A differentiation of eq.(\ref{eq:thermal}) gives:
\begin{multline} \label{eq:Cottrell4}
   \frac{dP(\Delta\sigma)}{d\Delta\sigma}=-\frac{V_a}{k_BT}P(\Delta\sigma)\\
   \Leftrightarrow P(\Delta\sigma)d\Delta\sigma=-\frac{k_BT}{V_a}dP(\Delta\sigma)
\end{multline}

Combining eqs.(\ref{eq:Cottrell4}, \ref{eq:Cottrell3} and \ref{eq:Cottrell1}) leads to:
\begin{multline} \label{eq:Cottrell5}
   \dot{\varepsilon}=\delta\epsilon \int_{0}^{1} n(\Delta\sigma,t=0)\text{exp}(-P(\Delta\sigma)t)\left(-\frac{k_BT}{V_a}\right) \,d{P(\Delta\sigma)}\\
   =\frac{1}{t}\frac{k_BT\delta\epsilon}{V_a} n(\Delta\sigma,t=0)\\
  \times \left(\text{exp}(-tP(\Delta\sigma=1))-\text{exp}(-tP(\Delta\sigma=0))\right)
\end{multline}

This can be reformulated as::
\begin{equation} \label{eq:Cottrell}
    \dot{\varepsilon}\sim\delta\epsilon\frac{ k_B T}{V_at}\left[e^{-t/t_m}-e^{-t/t_0}\right]
\end{equation}
, where $t_0=1/\nu_0$ and $t_m=t_0\text{exp}\left(\frac{V_a}{k_BT}\right)$. This expression predicts a logarithmic creep ($p$=1) between a small delay timescale $c=t_0$ and an upper timescale $t_m$ above which it vanishes, when all spots are exhausted. $\dot{\varepsilon}$ naturally increases with temperature, though not in an Arrhenius's way as proposed in eq.(\ref{eq:Carter}), while the $p$-value does not. In this oversimplified model, the applied stress is not explicitly considered and tertiary creep and failure cannot be modelled.

\section{Fiber bundle model with thermal activation} 

To release the crude assumption of the independence of events, we consider next a mean-field (democratic) load redistribution within a fiber-bundle model (FBM) with thermal activation. Several authors analyzed creep rupture within such a FBM framework by considering viscoelastic fibers that flow once they are broken \cite{kun2003scaling,jagla2011creep}, i.e. embedding thermally activated processes within the fibers rheology. Others superimposed a white thermal noise $\eta$ on the applied load $F$ \cite{roux2000thermally,ciliberto2001effect,politi2002failure}, but these authors did not focus on primary creep. We used a different (and numerically efficient) methodology, introducing a Kinetic Monte Carlo dynamics \cite{bortz1975new} within a democratic FBM. We consider an initial set of $N_0$ parallel fibers of constant elastic modulus $Y=1$ and of variable strength $\sigma_i\in[0,1]$, submitted to an applied load $F$ giving an initial fiber stress $\sigma(t=0)=\sigma_0=F/N_0$. The system is first suddenly loaded ($t$ remains fixed to 0) to $F$ in an athermal way. This implies that all fibers with a strength $\sigma_i\leq\sigma_0$ fail during this initial step, triggering a load redistribution, i.e. an increase of the fiber stress, so possibly additional fiber failures. For a uniform distribution of strengths, the critical stress $\sigma_c$ beyond which the bundle fails during this athermal loading is 1/4 \cite{pradhan2010failure}. 

\begin{figure}
\includegraphics[width=\linewidth]{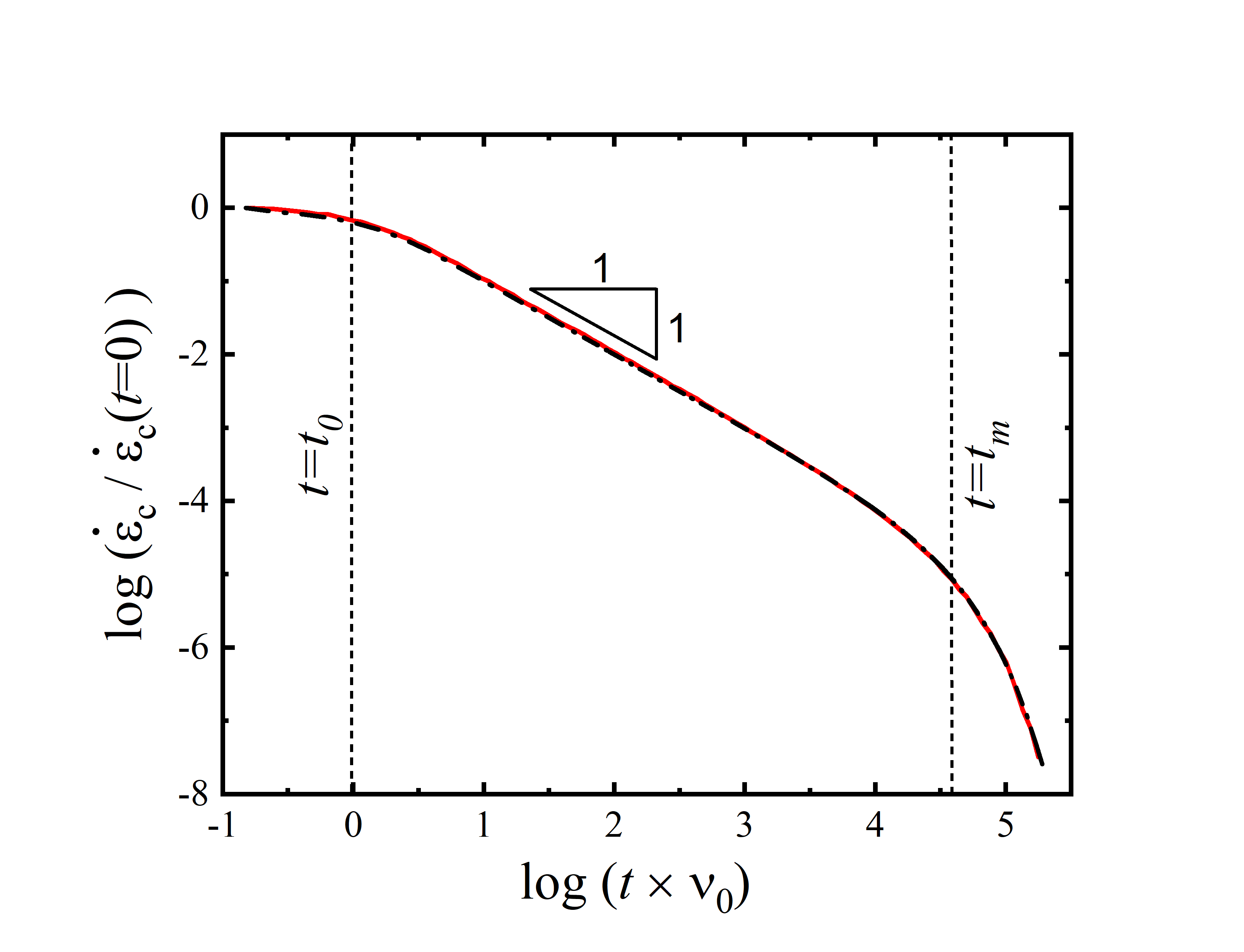}
\caption{\label{fig:FigureCottrell} Creep strain-rate: Comparison between eq.(\ref{eq:Cottrell}) (black dash-dotted line) and a “FBM” model without load sharing (red line).}
\end{figure}

At the end of this initial step, the number of remaining fibers is $N_1$, the fiber stress is $\sigma_1=F/N_1$ and the strain $\varepsilon_1=\sigma_1/Y$. Thermal activation is switched on from this point, maintaining $F<F_c=N_0\sigma_c$ constant. Deformation then occurs through a succession of thermally activated fiber ruptures, which are randomly selected according to the probabilities given by (\ref{eq:thermal}) with $\Delta\sigma_i(t)=\sigma_i-\sigma(t)$, meaning that the weakest spot (smallest $\Delta\sigma_i$) has a largest probability to fail, but is not necessarily selected systematically. The kinetic Monte-Carlo algorithm determines as well the time $\Delta t$ required for this event to happen. Such thermally activated ruptures redistribute the applied load, i.e. increase the effective fiber stress $\sigma(t)$, so potentially trigger \textit{athermal} avalanches of ruptures, as long as the fiber stress locally surpasses the strength of at least one element. We consider that the corresponding (short) elastic timescale is negligible compared to the (long) timescales of thermal activation and creep, so the clock is just stopped during these avalanches. Much like in the simple exhaustion model discussed above, a given fiber can only break once, and is then eliminated. Following \cite{herrmann1989fracture}, the fibers strengths are distributed according to $P(\sigma_i=x)=(1-b)x^{-b}$, where $b$ defines the strength of the disorder: $b=0$ corresponds to a uniform distribution, $b<0$ to a weak disorder and $0<b\lesssim1$ to a strong one. The results shown were obtained for $N_0=10^5$, averaging over 50 realizations of the disorder, but we checked that all our conclusions were independent of the system size.

We first note that, if the load sharing is switched off, i.e. the fiber stress remains constant ($\sigma(t)=\sigma_0$) and avalanches cannot occur, and setting $b=0$, this "FBM" model is equivalent to the crude exhaustion model presented above. We checked in this case the agreement between eq.(\ref{eq:Cottrell}) and our numerical results, see Fig. \ref{fig:FigureCottrell}.

\begin{figure}
\includegraphics[width=\linewidth]{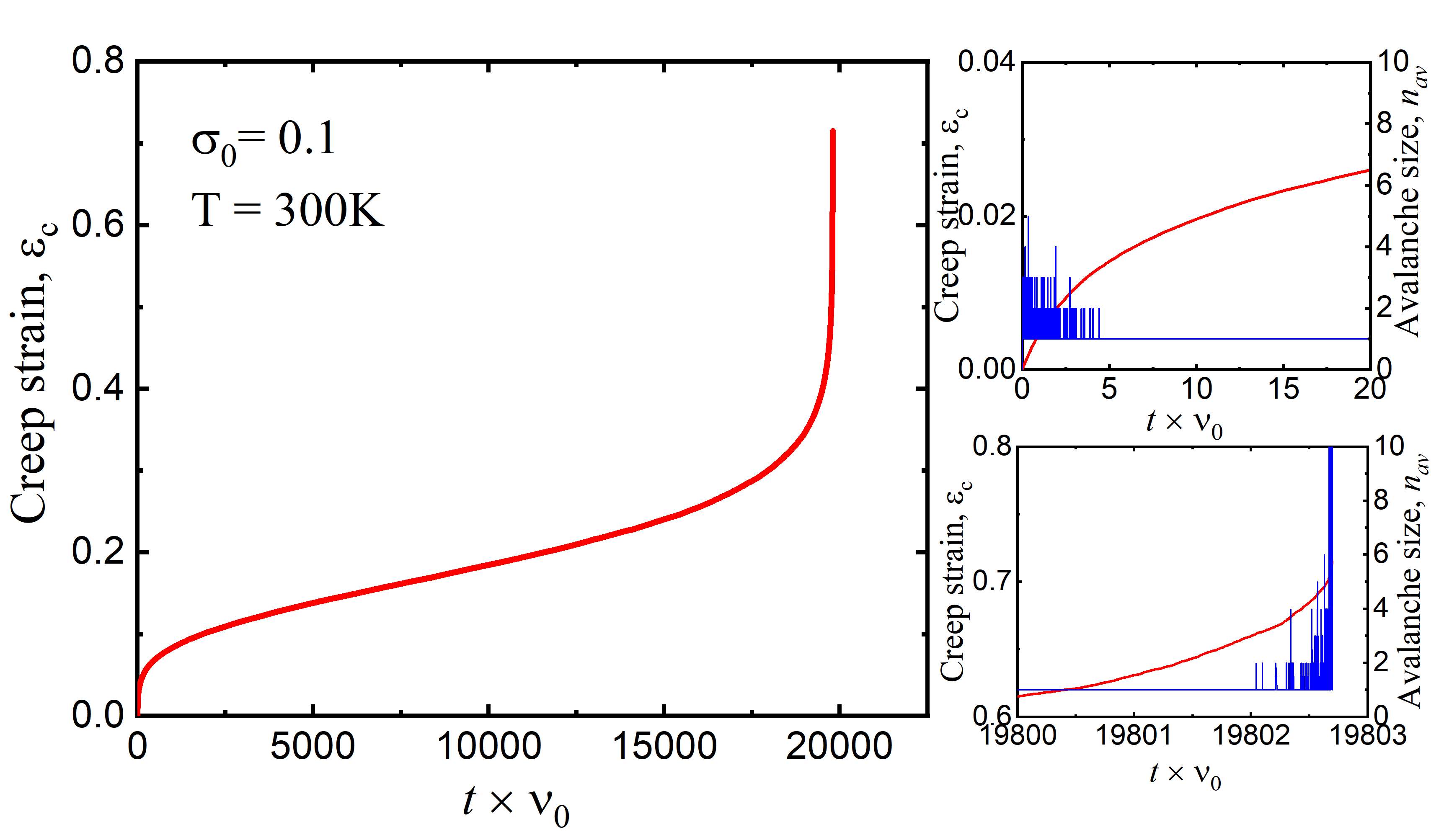}
\caption{\label{fig:FigCreep} Creep deformation from a thermally activated FBM showing (a) the classical phenomenology of creep ($\sigma_0$=0.1, $T$=300K, $b$=0). Zooms on the very early (b) and late (c) stages of creep experiencing some avalanche activity.}
\end{figure}

Instead, in the presence of load sharing, the model reproduces the entire phenomenology of creep deformation with a decelerating stage followed by an accelerating creep leading to system failure (Fig.~\ref{fig:FigCreep}a). Here we define the creep strain as $\varepsilon_c(t)=\varepsilon(t)-\varepsilon_1=(\sigma(t)-\sigma_1)/Y$, and the avalanche size $n_{av}$ as the number of fibers that break in a cascading sequence for each thermally activated event ($n_{av}\geq 1$). This avalanche activity qualitatively follows the evolution of creep deformation, decreasing rapidly during primary creep until fading away, i.e. $n_{av}\approx 1$ (Fig.~\ref{fig:FigCreep}b), before to accelerate before final failure (Fig.~\ref{fig:FigCreep}c). It also strongly depends on temperature or applied load: for low $T$ and/or $F$, athermal avalanches are nearly absent during primary creep, which therefore mainly results from thermally activated events (see below).

\begin{figure}
\includegraphics[width=\linewidth]{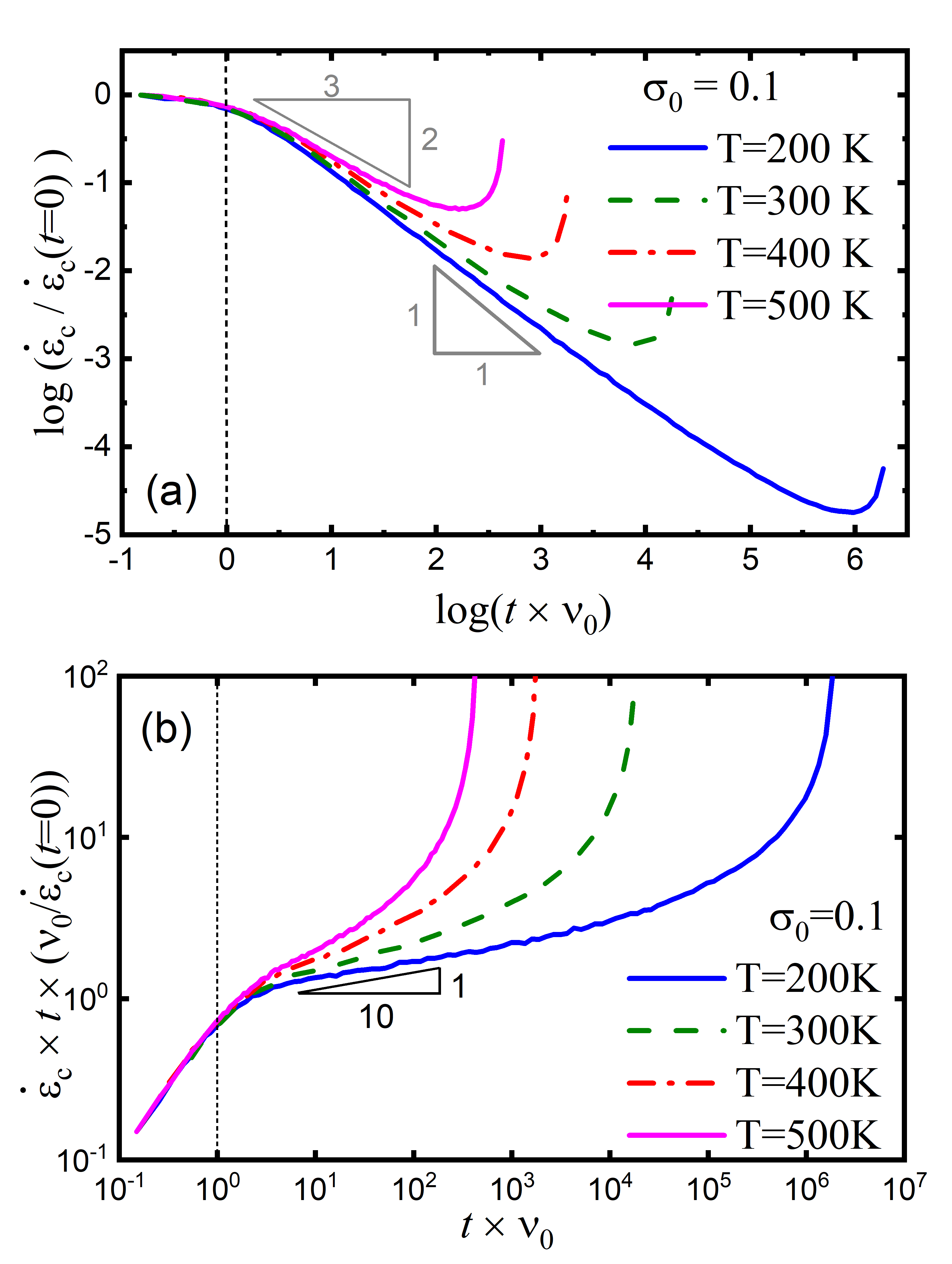}
\caption{\label{fig:FigureFBM1} (a) Effect of temperature $T$ on creep strain-rate $\dot{\varepsilon}_c$ in a thermally activated FBM. The slopes $p$=1 and $p$=2/3 are just given as a guide. (b) The same creep curves as in (a) multiplied by time $t$ to show that true logarithmic creep ($p$=1) is never observed, even at low temperature. At larger $T$ the apparent decrease of $p$ results from a transition between the initial plateau and the final acceleration during tertiary creep.}
\end{figure}

\begin{figure}
\includegraphics[width=\linewidth]{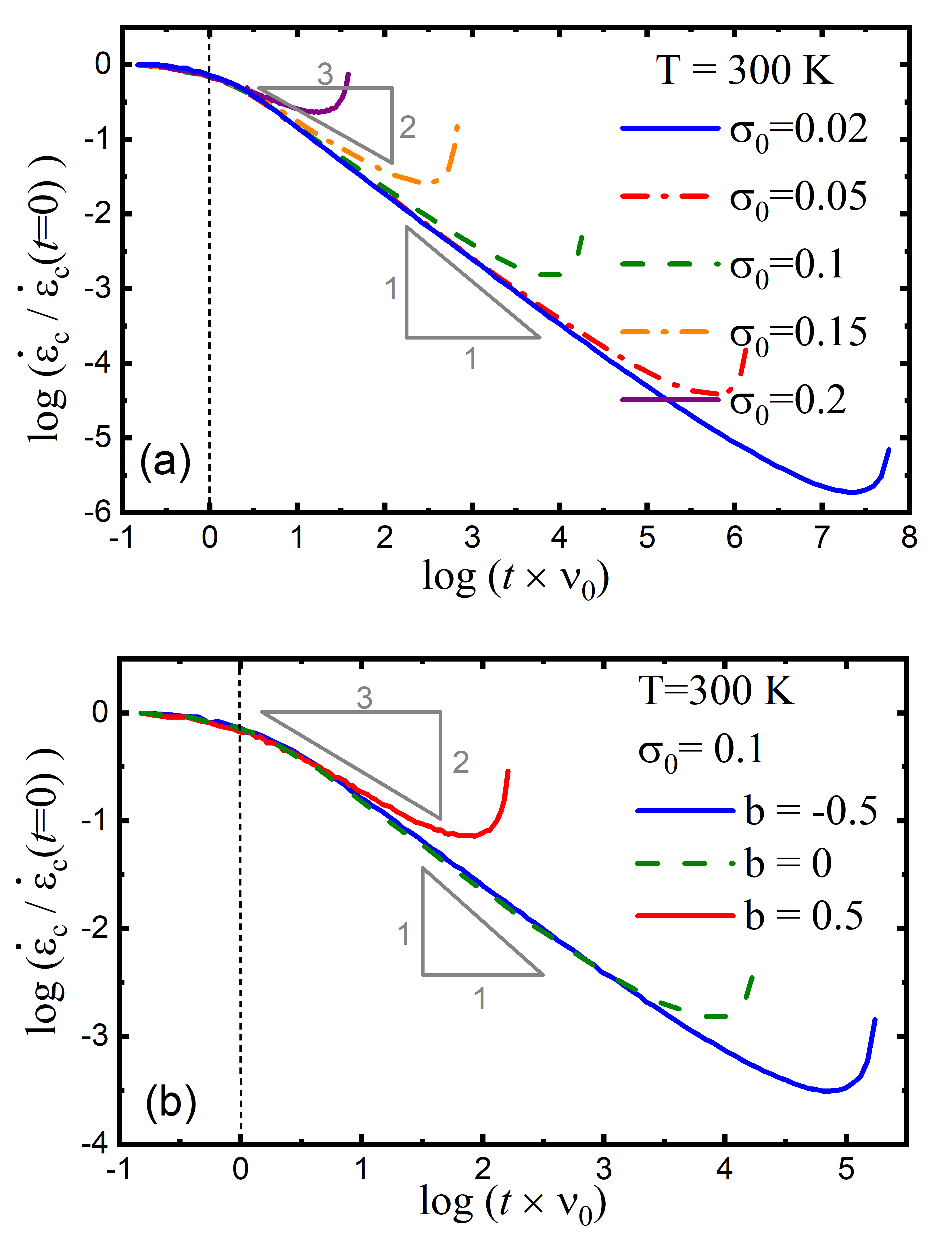}
\caption{\label{fig:FigureFBM2} Effect of (a) the applied stress $\sigma_0$ and (b) disorder strength $b$ on creep strain-rate $\dot{\varepsilon}_c$ in a thermally activated FBM. The slopes $p$=1 and $p$=2/3 are just given as a guide.}
\end{figure}

The introduction of load redistribution has several consequences. 

(i) The increasing number of broken fibers increases the effective stress $\sigma(t)=F/N_r(t)$ felt by the $N_r$ remaining ones, so decreases the stress gaps. This stimulates the creep dynamics from a shortening of the inter-events times $\Delta t$ between successive thermally activated events. 

(ii) In this FBM framework, this increasing effective stress also increases the strain increment $\delta\varepsilon$ produced by a single fiber rupture. Indeed, a differentiation of the creep strain $\varepsilon_c$ gives $\frac{d\varepsilon_c}{dN_r}=-\frac{F}{YN_r^2}$. Consequently, for a single fiber rupture, i.e. $N_r=-1$, the strain increment grows during creep as 
$\delta\varepsilon\sim1/N_r^2$. The conjunction of these two effects ((i) and (ii)) can be interpreted as a damage of the medium, at play even at very low $T$ or $\sigma_0$, when athermal avalanches are scarcely ever triggered. 

(iii) At higher temperatures and applied loads, such avalanche triggering sustains further the creep dynamics.  Altogether, these mechanisms imply that the creep strain-rate decreases more slowly during primary creep than for the oversimplified exhaustion model mentioned above, and eventually accelerates towards macrofailure. In other words, for the primary creep stage analyzed here, the interplay between thermal activation and load redistribution gives rise to a departure from logarithmic creep, and an apparent Andrade-like creep with $p<1$ . However, $p=2/3$ does not appear as a specific value. Instead, the apparent $p$-value continuously decreases with increasing temperature (Fig.~\ref{fig:FigureFBM1}a) and/or applied load  (Fig.~\ref{fig:FigureFBM2}a), in agreement with experimental data \cite{carter1978transient,cottrell1952time}. However, a more detailed analysis actually questions the existence of a genuine power law decay, at least at high $T$ or $\sigma_0$, while the dependence of $p$ on these parameters could result from a transition between an initial plateau of $\dot{\varepsilon}$ for $t<t_0$ and the acceleration preceding failure, which occurs earlier when $T$ or $\sigma_0$ increase (Fig.~\ref{fig:FigureFBM1}b). The dependence of the apparent $p$-value on temperature and the applied stress implies a failure of the empirical relation (\ref{eq:Carter}).  

\begin{figure}
\includegraphics[width=\linewidth]{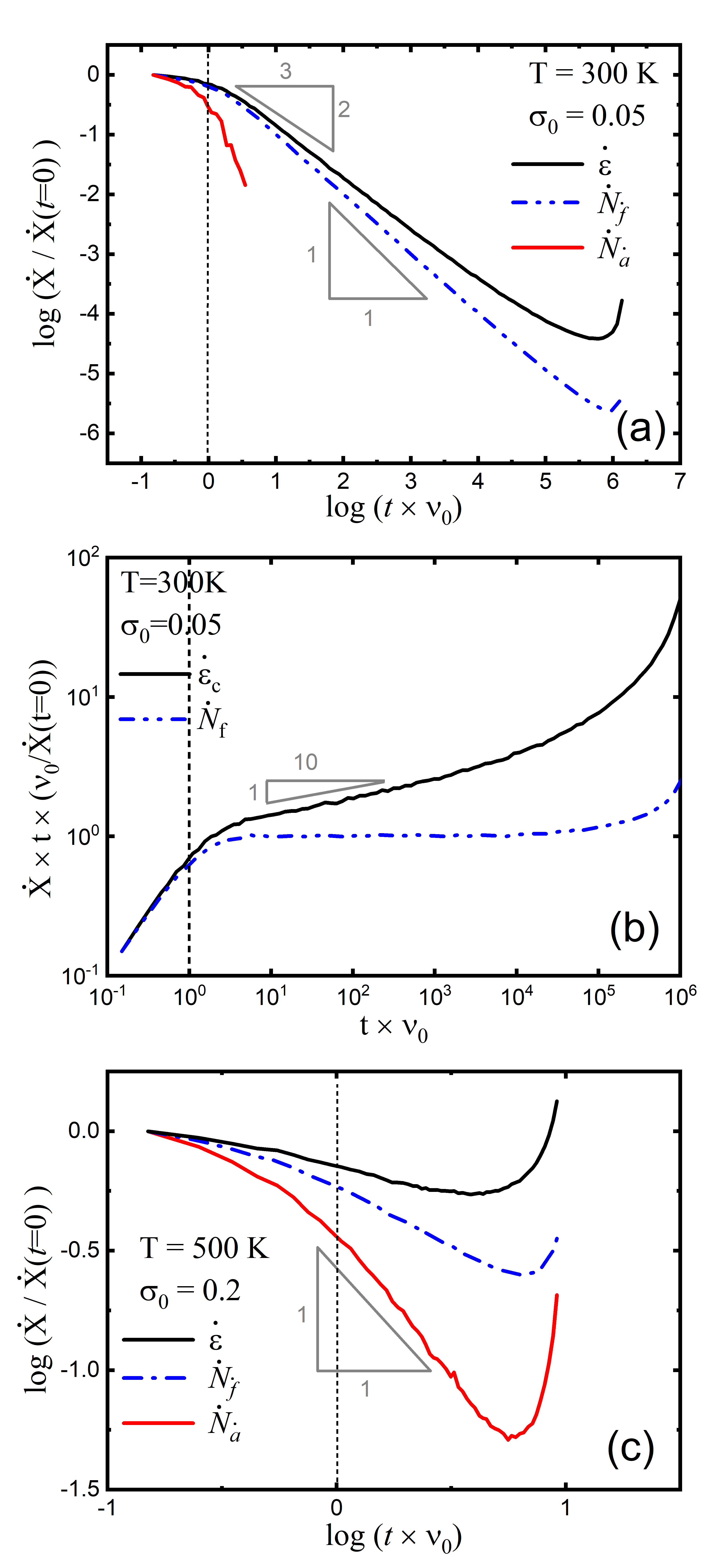}
\caption{\label{fig:FigureRates} (a) Evolution of the strain-rate $\dot\varepsilon_c$(black solid curve), the fiber breaking rate $\dot{N_f}$(blue dashed-dotted curve) and the athermal breaking rate $\dot{N_a}$(red solid curve) for a thermally activated FBM with relatively low $\sigma_0$ and $T$. (b) The strain-rate as well as the fiber breaking rate multiplied by $t$ in order to show the $1/t$ decay for $\dot{N_f}$ and the slower decay in $1/t^p$  with $p\simeq0.9$ for $\dot\varepsilon_c$.}
(c) Same as (a) for larger $\sigma_0$ and $T$. Uniform disorder ($b$=0) in both cases.
\end{figure}

The respective contributions of the mechanisms (i) to (iii) mentioned above can be analyzed from the evolution of the fiber breaking rate $\dot{N_f}$, where $N_f=N_1-N_r$, and of the athermal breaking rate $\dot{N_a}$, where $N_a$ is the number of athermally broken fibers (during avalanches). Tracking $N_f$ remains to consider that the strain increment $\delta\varepsilon$ remains constant during creep, i.e. to suppress point (ii). For low $\sigma_0$ and $T$, the avalanche activity is extremely limited, and vanishes rapidly for $t>t_0$ (Fig.~\ref{fig:FigCreep}b and ~\ref{fig:FigureRates}a), i.e. cannot explain the departure from logarithmic creep. Under these conditions, $\dot{N_f}\sim1/t$ over a significant range of timescales (Fig.~\ref{fig:FigureRates}a,b), while $p\simeq$0.9 (Fig.~\ref{fig:FigureFBM1}b and \ref{fig:FigureRates}b). This indicates that the impact of an increasing effective stress on thermal activation (point (i)) does not account alone for a departure from logarithmic creep, whereas the softening effect (point (ii)) induces a slight departure from the upper bound $p$=1. At larger loads and temperatures, both the damage of the system (points (i) and (ii)) and avalanche triggering (point (iii)) contribute to sustain the deformation during primary creep and therefore to decrease the apparent $p$-value. The evolution of the athermal breaking activity mimics a power law decay, $\dot{N}_a\sim(t+c)^{-p_a}$, though this merely results from a transition between an initial plateau for $t<t_c$ and a final acceleration before failure. Under these conditions, the total breaking rate $\dot{N}_f$ decays more slowly than $1/t$, meaning that the shortening of inter-events times as the result of a larger effective stress impacts significantly the apparent $p$-value (Fig.\ref{fig:FigureRates}c).

Much like for increasing temperature or applied load, an increasing disorder strength decreases creep lifetimes (Fig.~\ref{fig:FigureFBM2}b). This is expected, as this implies more fibers with small stress gaps $\Delta\sigma$, even at the onset of creep loading, after the athermal pre-loading. For weak ($b<0$) to uniform ($b=0$) disorder, this does not seem to have a large impact on the $p$-value. For stronger disorder ($1>b>0$), the reduction of the duration of the primary creep stage is strong enough to lead to a significant decrease of the apparent $p$-value. These results are consistent with the effect of aging observed in colloidal glasses \cite{siebenburger2012creep}, which is expecting to eliminate the weakest spots. 

\begin{figure}
\includegraphics[width=\linewidth]{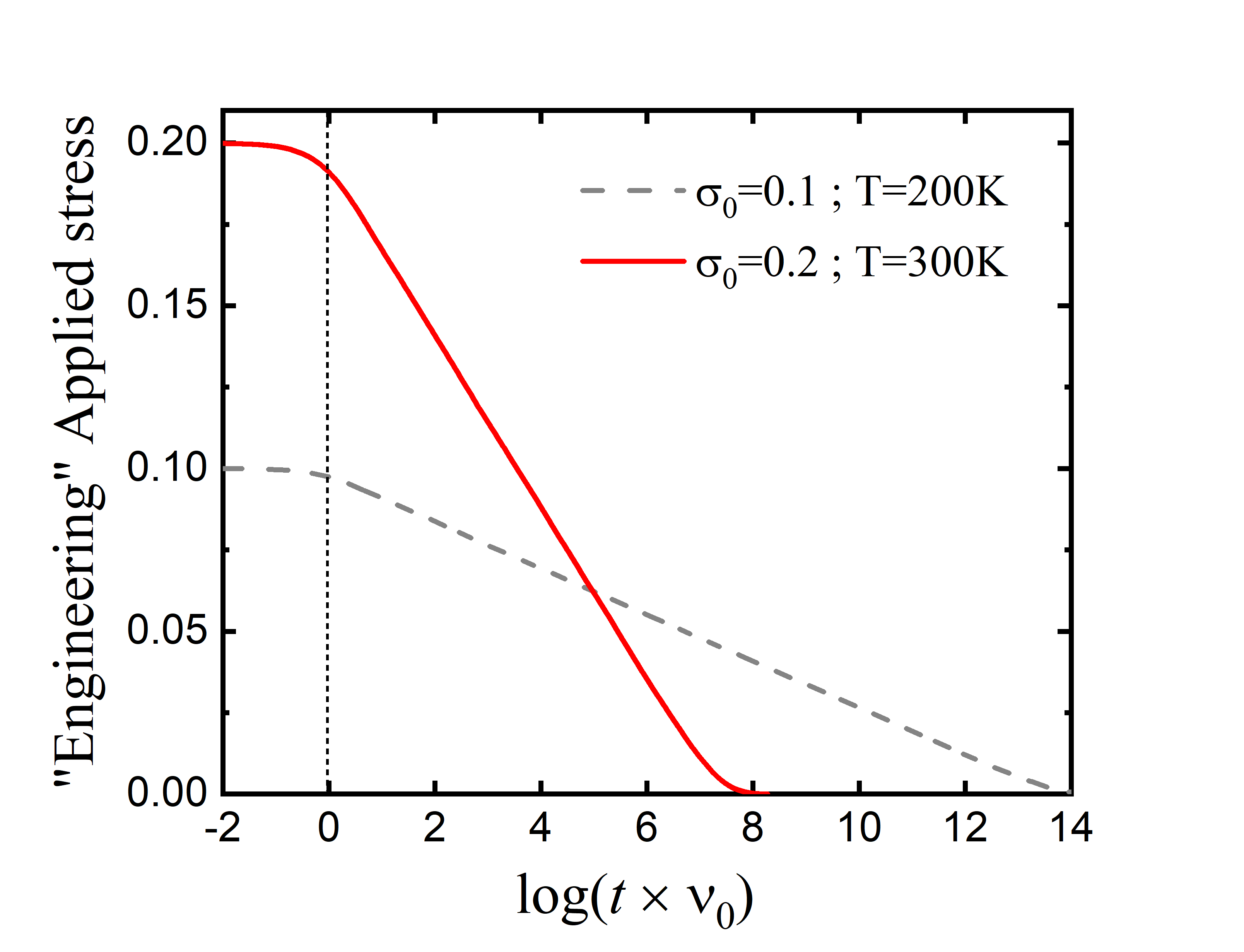}
\caption{\label{fig:FigRelaxation} Relaxation of the applied stress under a constant imposed strain $\varepsilon_0$ for a thermally activated democratic FBM. A logarithmic relaxation is observed whatever the initial engineering stress $\sigma_0=F_0/N_0$ and/or temperature $T$.}
\end{figure}

The same democratic FBM can be considered under relaxation conditions. This means increasing athermally the initial load up to $F(t=0)=F_0$, hence triggering some fiber breaking, then maintaining the strain $\varepsilon=\varepsilon_0$ constant. During the relaxation stage, the effective stress $\sigma=\frac{F}{N}$ felt by the unbroken fibers remains strictly constant, $\sigma=Y\varepsilon_0$. Consequently, each individual fiber breaking leads to a load drop $\frac{dF}{dN}=-Y\varepsilon_0$. Under these conditions, athermal avalanches are absent and the thermally activated exhaustion mechanism leads to $\dot{N}_f\sim\frac{1}{t}$, much like what is observed under creep conditions for low applied load and temperature (see Fig.~\ref{fig:FigureRates}b). This leads to a logarithmic relaxation of the applied load, $F(t)\sim-\text{log}(t+c)$, or of the "engineering" applied stress $F/N_0$, as $\frac{dF}{dt}=\frac{dF}{dN_f}\frac{dN_f}{dt} \sim \frac{-Y\varepsilon_0}{t}$. Numerical simulations are fully consistent with this scenario, whatever $F_0$ and/or $T$ (Fig.~\ref{fig:FigRelaxation}). In other words, the relaxation exponent is always equal to 1, and there is no equivalent of Andrade's creep for this loading mode. Such behavior is reminiscent of several other physical systems characterized by slow, logarithmic relaxations, such as conductance relaxation in electron glasses \cite{amir2012relaxations}, magnetic relaxation in superconductors \cite{kodama1997relation}, or vibration-induced compaction in granular media \cite{josserand2000memory}, raising interesting potential connections.

\section{Progressive damage model} 

\begin{figure}
\includegraphics[width=\linewidth]{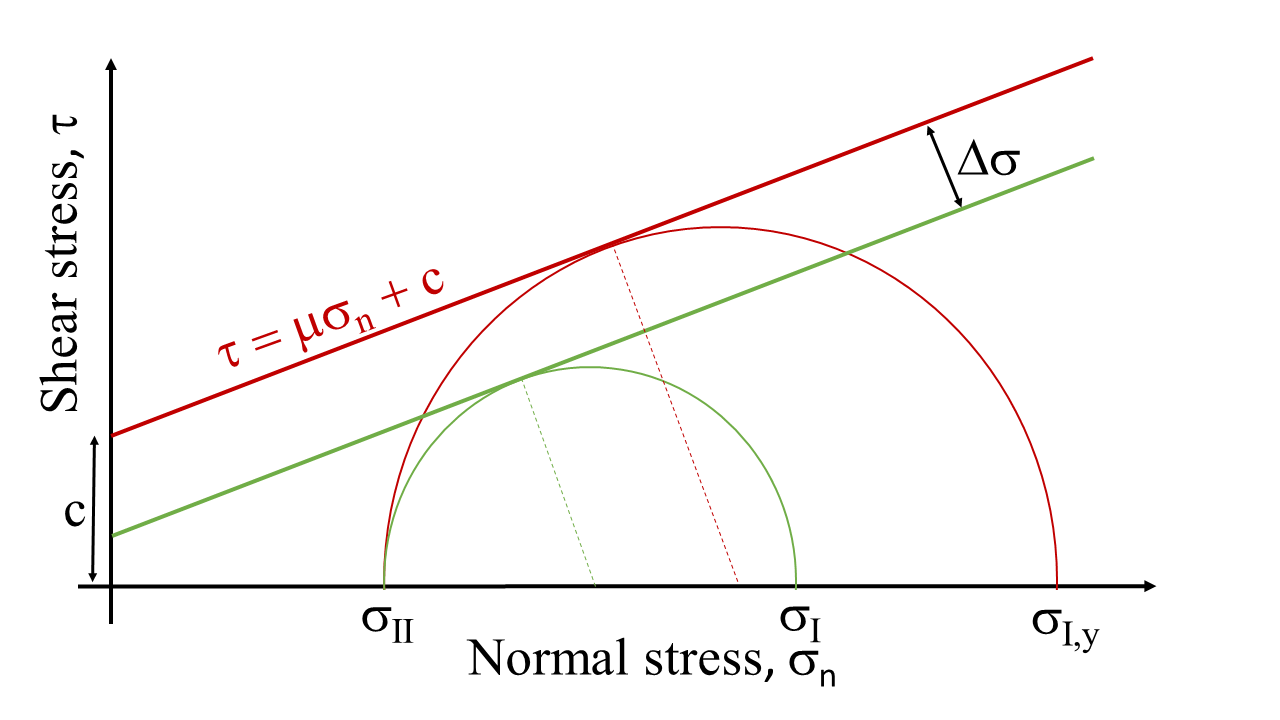}
\caption{\label{fig:FigCoulomb} Coulomb's damage criterion used in the PDM within a Mohr-Coulomb diagram, and the associated definition of the stress gap $\Delta\sigma$ used for the thermal activation of damage events (eq. \ref{eq:thermal}). The red semi-circle represents the state of stress allowing an athermal damage for a cohesion $c$ and an internal friction $\mu$, while the green semi-circle represents a sub-critical state of stress requiring thermal activation to trigger damage.}
\end{figure}

\begin{figure}
\includegraphics[width=\linewidth]{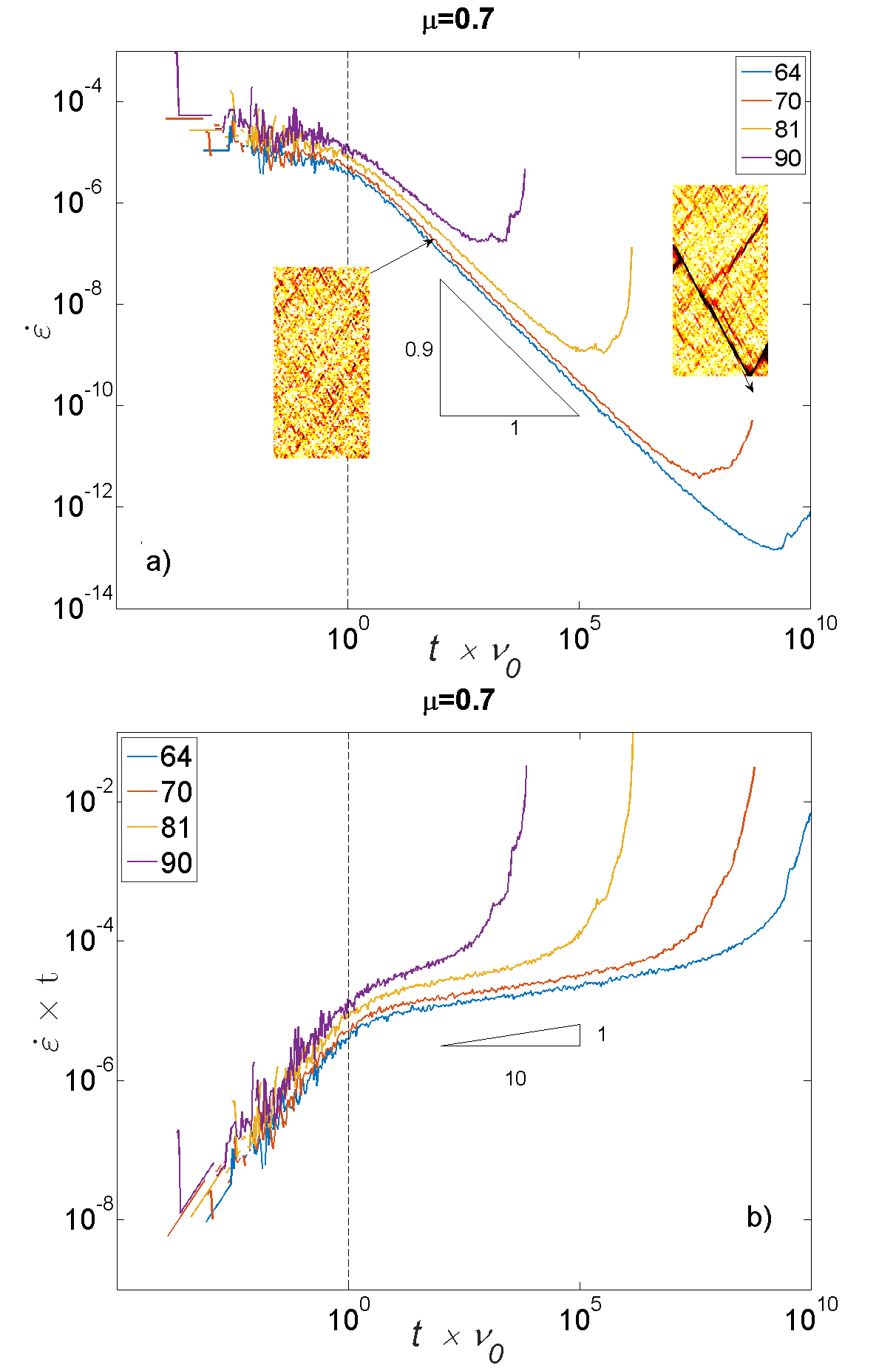}
\caption{\label{fig:FigurePGM07} (a) Evolution of the strain-rate $\dot\varepsilon_c$ for a thermally activated progressive damage model ($T$=600 K) with $\mu$=0.7. The different creep curves, averaged over 10 realizations of the disorder for system sizes 16$\times$32 elements, correspond to applied uniaxial stresses varying between 64\% and 90\% of the athermal failure stress for an identical initial microstructure. The insets show the damage field for an individual simulation at 70\% of the athermal strength, during primary creep (left) and close to failure (right). (b) The same creep curves as in (a) multiplied by $t$ to show that true logarithmic creep ($p$=1) is never observed.}
\end{figure}

\begin{figure}
\includegraphics[width=\linewidth]{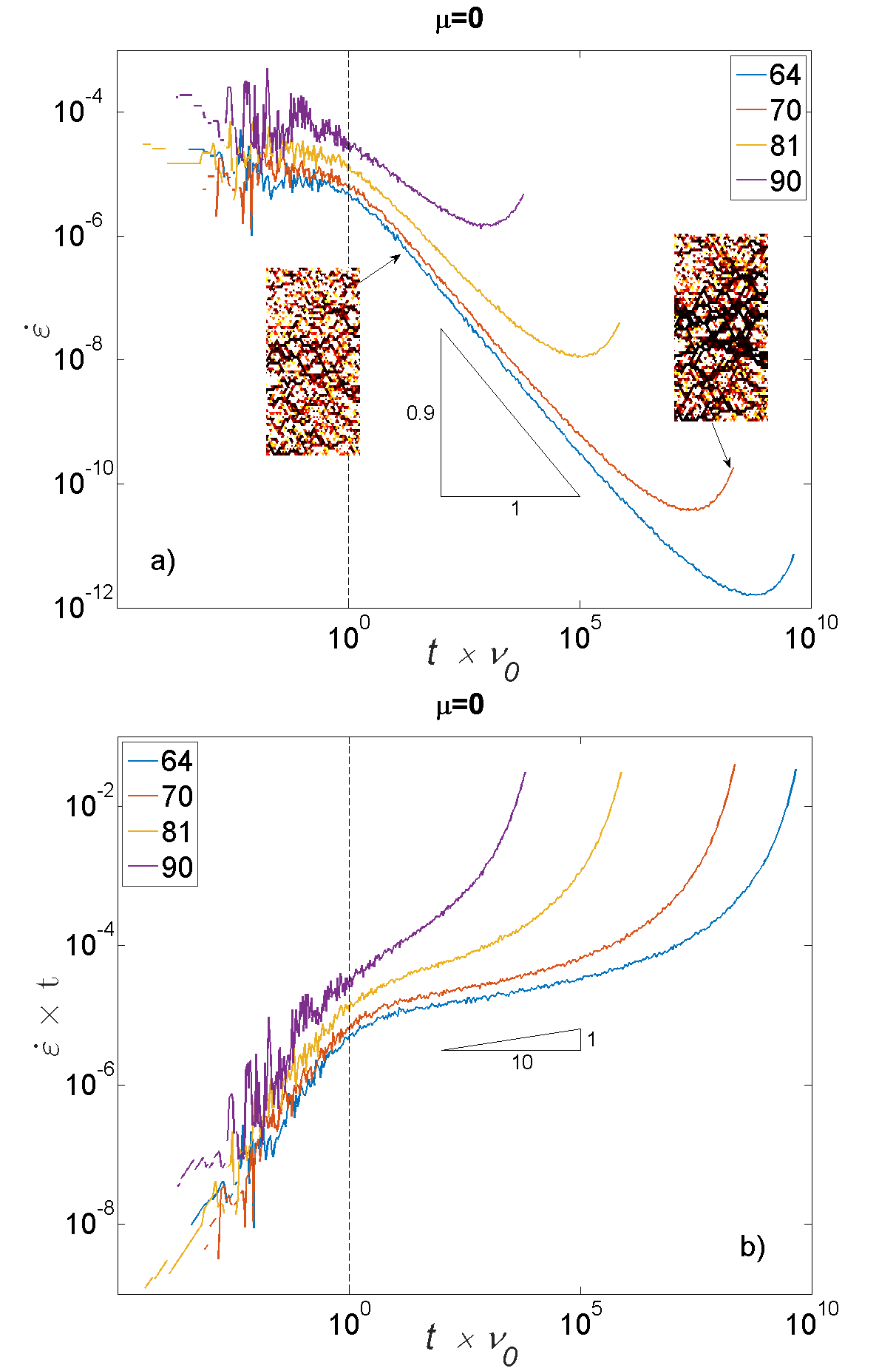}
\caption{\label{fig:FigurePGM00} Same as Fig.~\ref{fig:FigurePGM07} for $\mu$=0.}
\end{figure}

The mean-field (democratic) FBM framework considered so far remains crude, especially in terms of the elastic redistribution kernel which, in elastic solids, is non-convex and decays with distance as a power law \cite{eshelby1957determination}. The exhaustion hypothesis, i.e. the fact that a local site cannot deform more than once, represents another simplification. 
To evaluate the potential impact of these shortcomings, we introduced a similar Kinetic Monte-Carlo algorithm within a progressive damage model (PDM). The athermal version of this PDM has been extensively detailed elsewhere \cite{amitrano1999diffuse,girard2010failure}, and we recall its main features here. We consider a continuous 2D isotropic elastic domain (Hooke's law) under plane stress, modelled using a finite element scheme. Progressive local damage is represented by a decrease of the Young's modulus $Y_i$ of element $i$, $Y_i(n+1)=Y_i(n)d_0$ (with $d_0$=0.9) each time the stress state on that element exceeds a threshold, while the Poisson's ratio $\nu$ remains unchanged, and constant over all the elements. This procedure mimics, at the meso/element-scale, the elastic softening resulting from microcracking and damage \cite{kachanov1958time}. 
The simulations start with a homogeneously elastic, undamaged material, $Y_i=Y_0$ $\forall i$. The damage threshold is defined from the Coulomb's criterion, $\tau=\mu\sigma_N+c$ where $\tau$ and $\sigma_N$ are respectively the shear and normal stress over the orientation that maximizes the Coulomb's stress $\tau-\mu\sigma_N$ on the element (positive sign convention in compression; see Fig.~\ref{fig:FigCoulomb}), where $\mu$ is an internal friction coefficient, constant in space and time. We performed simulations for $\mu$=0.7 (Fig.~\ref{fig:FigurePGM07}), a common value for geomaterials \cite{jaeger2009fundamentals}, as well as for $\mu$=0 (Fig.~\ref{fig:FigurePGM00}), corresponding to a Tresca's criterion with no more impact of the normal stress on the threshold.

Quenched disorder is introduced from a uniform distribution of the cohesion, $0.15\times10^{-3}Y_0\leq c\leq 1.5\times10^{-3}Y_0$. Uniaxial compression is applied to the system, and the internal stress field is recalculated each time a damage event occurs by solving the static equilibrium, using the finite-element scheme. This way, a non-convex elastic redistribution kernel is naturally reproduced after a local damage event \cite{dansereau2019collective}. Such stress redistribution implies that the \emph{local} stress field is not necessarily uniaxial anymore, i.e. the minimum principal stress $\sigma_{II} \neq 0$ at the element scale.
Under monotonic loading, this athermal model was shown to successfully reproduce the main characteristics of Coulombic failure in disordered materials, such as the progressive localization of damage upon approaching a peak stress at which an incipient fault nucleates, or the impact of confining pressure and of the internal friction on strength and brittleness \cite{amitrano1999diffuse, amitrano2003brittle}. Here, thermally activated damage is introduced at the element scale from eq.(\ref{eq:thermal}) using a Kinetic Monte-Carlo scheme, with $\Delta \sigma$ the Coulomb's stress gap (see Fig.~\ref{fig:FigCoulomb}), and the system is loaded under a constant compressive external ("engineering") stress representing a fraction of the failure stress obtained for athermal monotonic loading. Much like for the FBM, the phenomenology of creep deformation and rupture is recovered, accompanied here by a progressive localization of damage during tertiary creep, a consequence of the non-convex kernel, see Fig.~\ref{fig:FigurePGM07} for an internal friction $\mu$=0.7. Here we focus on primary creep, for which extended localization is not observed. At low applied stresses, a robust power law decay of $\dot\varepsilon_c$ is observed over a large range of timescales with $p\simeq$0.9, similarly to FBM. This behavior is observed independently of the internal friction $\mu$ (see Fig.~\ref{fig:FigurePGM00} for $\mu$=0). At larger stresses, close to the athermal strength, the range of timescales over which an apparent powerlaw decay of $\dot{\varepsilon}_c$ is observed shrinks significantly, with a smaller apparent $p$-value. Much like for the FBM, this is the consequence of an increasing role of elastic stress redistribution on primary creep dynamics. Although the damage patterns depend, as expected, on the Coulomb's stress redistribution kernel and so on the internal friction $\mu$, the primary creep phenomenology does not (compare Fig.~\ref{fig:FigurePGM07} and Fig.~\ref{fig:FigurePGM00}).

Overall, the agreement between PDM and FBM indicates that the value of the internal friction $\mu$, the nature of the elastic redistribution kernel, or a strict exhaustion of weak spots have a little effect on the nature of primary creep. It is consistent with limited damage localization during this stage. This argues for the universal character of the scenario proposed here.   

\section{Summary and conclusions} We conclude that logarithmic primary creep appears as an upper bound for the $p$-value, associated to situations where stress redistribution and elastic interactions do not play a significant role, and deformation results from the accumulation of thermally activated, independent small events. Instead, stress redistribution implies that the occurrence of deformation events depends on the previous history. This shortens the time between successive events, increases their average amplitude, and potentially triggers deformation avalanches. All of this sustains creep deformation, leading to lower apparent $p$-values. In cases where damage avalanches are nearly nonexistent during primary creep (low $T$ and/or $F$), a $p\simeq$0.9 value seems to emerge, a value consistent with experimental data on concrete \cite{makinen2022}. 
We considered here a "brittle creep" framework where deformation results from rupture or damage events, while Andrade-like creep has been observed as well in systems where the microscopic deformation mechanisms are not supposed to damage the material, e.g. dislocation-driven creep in metals. However, Cottrell \cite{cottrell2004microscopic} proposed in this case that thermal activation and stress redistribution could indeed combine to trigger dislocation avalanches and Andrade's creep. We therefore suggest that our scenario is not restricted to brittle creep, but relevant as well to any type of local deformation mechanism, as long as it is thermally activated and leads to long-range stress redistribution.
On the other hand, in case of dislocation-driven creep, an alternative explanation based only on collective dislocation interactions and topological constrains, without invoking thermal activation, has been proposed as well \cite{miguel2002dislocation}. A possible extension to other systems such as glasses \cite{siebenburger2012creep} or soft matter \cite{divoux2011stress} would be also worth exploring in the future. Under relaxation conditions (strain, not stress, maintained constant), the same scenario leads to a logarithmic relaxation of the load, mirroring logarithmic creep under constant stress, but there is no equivalent of Andrade-like creep in that case. Such slow relaxation raises interesting comparisons with entirely different physical systems such as electron glasses, superconductors, or granular media.

\begin{acknowledgments}
{\em Acknowledgments -} 
ISTerre is part of Labex OSUG@2020. We thank two anonymous reviewers for interesting suggestions on this manuscript, and J.C. Verano Espitia for running some of the PDM simulations.
\end{acknowledgments}

\bibliography{biblio}

\end{document}